\pdfoutput = 1
\documentclass[11pt]{article}
\usepackage{url}
\usepackage{graphicx}
\usepackage{amsfonts}
\usepackage{amssymb}
\usepackage{latexsym}
\usepackage{color}
\usepackage{enumerate}
\usepackage{hyperref}

\newcommand{\notyet}[1]{}

\newcommand{\ABox}{
\raisebox{3pt}{\framebox[6pt]{\rule{6pt}{0pt}}}
}

{\makeatletter
 \gdef\xxxmark{%
   \expandafter\ifx\csname @mpargs\endcsname\relax 
     \expandafter\ifx\csname @captype\endcsname\relax 
       \marginpar{xxx}
     \else
       xxx 
     \fi
   \else
     xxx 
   \fi}
 \gdef\xxx{\@ifnextchar[\xxx@lab\xxx@nolab}
 \long\gdef\xxx@lab[#1]#2{{\bf [\xxxmark #2 ---{\sc #1}]}}
 \long\gdef\xxx@nolab#1{{\bf [\xxxmark #1]}}
 \gdef\turnoffxxx{\long\gdef\xxx@lab[##1]##2{}\long\gdef\xxx@nolab##1{}}%
}

\def\nothing{\ast}

\begin{document}

\title{Unfolding Orthogonal Polyhedra with Quadratic Refinement:\\
The Delta-Unfolding Algorithm}
\author{Mirela Damian%
   \thanks{Dept. of Computing Sciences, Villanova University, Villanova,
    PA 19085, USA.
   \protect\url{mirela.damian@villanova.edu}.}
\and
Erik D. Demaine%
     \thanks{Computer Science and Artificial Intelligence Laboratory,
       Massachusetts Institute of Technology,
       32 Vassar St., Cambridge, MA 02139, USA. \protect\url{edemaine@mit.edu}.
       Partially supported by NSF CAREER award CCF-0347776.}
\and
Robin Flatland%
   \thanks{Dept. of Computer Science, Siena College, Loudonville, NY 12211, USA.
    \protect\url{flatland@siena.edu}.}
}

\date{}
\maketitle

\begin{abstract}
  We show that every orthogonal polyhedron homeomorphic to a sphere can be
  unfolded without overlap while using only polynomially many (orthogonal) cuts.
  By contrast, the best previous such result used exponentially many cuts.
  More precisely, given an orthogonal polyhedron with $n$ vertices,
  the algorithm cuts the polyhedron only where it is met by
  the grid of coordinate planes passing through the vertices,
  together with $\Theta(n^2)$ additional coordinate planes
  between every two such grid planes.
\end{abstract}

\section{Introduction}

One of the major unsolved problems in geometric folding is whether every
polyhedron (homeomorphic to a sphere) has an ``unfolding''
\cite{Bern-Demaine-Eppstein-Kuo-Mantler-Snoeyink-2003,Demaine-O'Rourke-2007}.
In general, an \emph{unfolding} consists of cutting along the polyhedron's
surface such that what remains flattens into the plane without overlap.
Convex polyhedra have been known to unfold
since at least the 1980s \cite[Sec.~24.1.1]{Demaine-O'Rourke-2007}.

A recent breakthrough for nonconvex polyhedra is the unfolding of
any ``orthogonal'' polyhedron (homeomorphic to a sphere)
\cite{Damian-Flatland-O'Rourke-2007-epsilon}.
A polyhedron is \emph{orthogonal} if all of its edges are parallel
to a coordinate axis, and thus all edges and faces meet at right angles.
While very general, a disadvantage of this unfolding algorithm
is that the cutting is inefficient,
making exponentially many cuts in the worst case, 
resulting in an unfolding that is long and thin (``epsilon thin'').
         
In this paper, we show how to unfold any orthogonal polyhedron
using only a polynomial number of cuts.

\paragraph{Grid refinement.}
To more precisely quantify the cuts required by an unfolding,
several models of allowed cuts have been proposed.
See \cite{Demaine-O'Rourke-2005,Demaine-O'Rourke-2007,O'Rourke-2008-orthosurvey}
for surveys.

For convex polyhedra, the major unsolved goal is to just cut along the edges
(which implies a linear number of cuts) \cite[ch.~22]{Demaine-O'Rourke-2007}.
For nonconvex polyhedra, however, this goal is unattainable, even
when the polyhedron is ``topologically convex''
\cite{Bern-Demaine-Eppstein-Kuo-Mantler-Snoeyink-2003}
or is orthogonal
\cite{Biedl-Demaine-Demaine-Lubiw-Overmars-O'Rourke-Robbins-Whitesides-1998}.
A simple example of the latter is a small box on top of a larger box.
More generally, deciding whether an orthogonal polyhedron has an edge
unfolding is strongly NP-complete \cite{Abel-Demaine-2011}.

For orthogonal polyhedra, it seems most natural to consider orthogonal cuts.
The smallest extension from edge unfolding seems to be \emph{grid unfolding}
(a concept implicit in
\cite{Biedl-Demaine-Demaine-Lubiw-Overmars-O'Rourke-Robbins-Whitesides-1998}),
where we slice the polyhedron with all axis-aligned planes that pass through
at least one polyhedron vertex, and allow cutting along all slice lines.
Even with these additional edges, few nontrivial subclasses of
orthogonal polyhedra are known to have grid unfoldings: ``orthotubes''
\cite{Biedl-Demaine-Demaine-Lubiw-Overmars-O'Rourke-Robbins-Whitesides-1998},
``orthostacks'' composed of orthogonally convex slabs
\cite{Damian-Meijer-2004-orthostacks}, and
``well-separated orthotrees''
\cite{Damian-Flatland-Meijer-O'Rourke-2005-orthotrees}.
On the negative side, there are four orthogonal polyhedra with no
\emph{common} grid unfolding
\cite{Aloupis-Bose-Collette-Demaine-Demaine-Douieb-Dujmovic-Iacono-Langerman-Morin-2010}.

The next extension beyond grid unfolding is \emph{grid refinement} $k$,
which additionally slices with $k$ planes in between every grid plane
(as above), and allows cuts along any edges of the refined grid.
With constant grid refinement, a few more classes of orthogonal polyhedra
have been successfully unfolded: orthostacks
\cite{Biedl-Demaine-Demaine-Lubiw-Overmars-O'Rourke-Robbins-Whitesides-1998},
and Manhattan towers \cite{Damian-Flatland-O'Rourke-2008-manhattan}.

The breakthrough was the discovery that arbitrary orthogonal polyhedra
(homeomorphic to a sphere) unfold with finite grid refinement
\cite{Damian-Flatland-O'Rourke-2007-epsilon}.
Unfortunately, the amount of grid refinement is exponential
in the worst case (though polynomial for ``well-balanced'' polyhedra).
For this reason, the unfolding algorithm was called
\emph{epsilon-unfolding}.

\paragraph{Our results.}
We show how to modify the epsilon-unfolding algorithm of
\cite{Damian-Flatland-O'Rourke-2007-epsilon}
to reduce the refinement from worst-case exponential ($2^{\Theta(n)}$)
to worst-case quadratic ($\Theta(n^2)$), while still unfolding
any orthogonal polyhedron (with $n$ vertices) homeomorphic to a sphere.
We call our algorithm the \emph{delta-unfolding} algorithm,
to suggest that the resulting surface strips are still narrow
but wider than those produced by epsilon-unfolding.

Our central new technique in delta-unfolding is the concept of ``heavy''
and ``light'' nodes from ``heavy-path decomposition''
\cite{Sleator-Tarjan-1983}.  Interestingly, heavy-path decomposition
is a common technique for balancing trees in the field of data structures,
but not so well known in computational geometry.

Even with this technique in hand, however, delta-unfolding requires a
careful modification and engineering of the techniques used by
epsilon-unfolding.  Thus, Sections~\ref{sec:overview}
and~\ref{Epsilon-Unfolding Extrusions} start with reviewing the
main techniques of epsilon-unfolding; then Section~\ref{sec:Delta}
modifies those techniques; and finally
Section~\ref{Delta-Unfolding of Genus-Zero Orthogonal Polyhedra}
puts these techniques together to obtain our main result.

\section{Overview of Epsilon-Unfolding}
\label{sec:overview}

We begin with a review the epsilon-unfolding algorithm~\cite{Damian-Flatland-O'Rourke-2007-epsilon},
starting in this section with a high-level overview,
and then in Section~\ref{Epsilon-Unfolding Extrusions}
detailing those aspects of the algorithm
that we modify to achieve quadratic refinement.

Throughout this paper,
$P$ denotes a genus-zero orthogonal polyhedron whose edges are parallel to the
coordinate axes and whose surface is a $2$-manifold.
We take the $z$-axis to define the \emph{vertical} direction,
the $x$-axis to determine \emph{left} and \emph{right},
and the $y$-axis to determine \emph{front} and \emph{back}.
%
%
We consistently take the viewpoint from $y=-\infty$.
The faces of $P$ are distinguished by their outward normal:
forward is $-y$; rearward is $+y$; left is $-x$; right is $+x$; bottom is $-z$;
top is $+z$.%
\footnote{The $\pm y$ faces are given the awkward names
  ``forward'' and ``rearward'' to avoid confusion with other uses
  of ``front'' and ``back'' introduced later.}

The epsilon-unfolding algorithm partitions $P$ into \emph{slabs}
by slicing
it with $y$-perpendicular planes through each vertex. Let $Y_0, Y_1, Y_2, \dots$
be the slicing planes sorted by $y$ coordinate. A \emph{slab} 
$s$ is a connected component of
$P$ located between two consecutive planes $Y_i$ and $Y_{i+1}$.
Each slab is a simple orthogonal polygon extruded in the $y$-direction.
The cycle of
\{left, right, top, bottom\} faces surrounding $s$ is called a \emph{band}, and
the band edges in $Y_i$ (and similarly in $Y_{i+1}$) form a cycle called a \emph{rim}.
A $z$-\emph{beam} is a narrow vertical strip on a forward or rearward face of $P$
connecting the rims of two bands.
The order in which the bands unfold is determined
by an \emph{unfolding tree} $T_U$ whose nodes are bands,
and whose arcs correspond to $z$-beams, each of which connects a
parent band to a child band in $T_U$. The unfolding tree
$T_U$ will be further described in Section~\ref{sec:Tree} below.

The unfolding of a band $b$ is determined by
a thin surface spiral, 
denoted $\xi$, that starts on one of $b$'s rims,
cycles around $b$ 
while displacing toward the other rim, where it
turns around and returns to the point
it started.  As the spiral passes by a $z$-beam connecting $b$ to one of its children, 
it enters through the $z$-beam to the child's rim, and then recursively
visits the subtree rooted at the child.
Once the complete spiral 
is determined, it can be thickened in the $\pm y$ direction so that
it entirely covers all band faces.
The thickened spiral 
is such that it can be laid flat in the plane to form a
monotonic staircase strip. The forward and rearward faces of $P$ can then be laid flat without overlap
by attaching them in strips above and below the staircase.

\section{Epsilon-Unfolding Extrusions}
\label{Epsilon-Unfolding Extrusions}

Almost all algorithmic issues in epsilon-unfolding are present in unfolding polyhedra
that are $z$-extrusions of simple orthogonal polygons in the $xy$ plane.
Therefore, we follow~\cite{Damian-Flatland-O'Rourke-2007-epsilon} in describing
the algorithm for this simple shape class, before extending the ideas
to all orthogonal polyhedra.  All modifications needed for
delta-unfolding
are also present in unfolding orthogonal extrusions, and so we
describe them in terms of this simple shape class.
We therefore review in detail the epsilon-unfolding algorithm for orthogonal extrusions.

\subsection{Unfolding Tree}
\label{sec:Tree}
\begin{figure}[htbp]
\centering
\includegraphics[width=.7\linewidth]{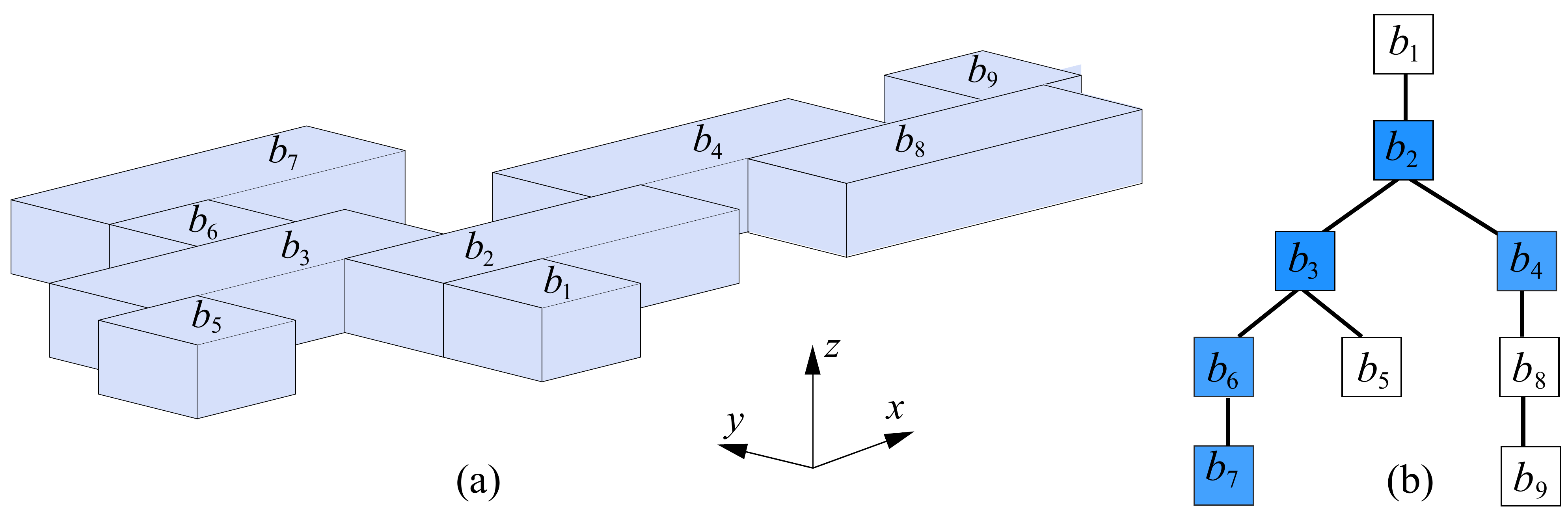}
\caption{(a) Extrusion of an orthogonal polygon, partitioned by $y$ perpendicular planes. (b) Unfolding
  tree. Back children are represented by shaded nodes.}
\label{fig:partition}
\end{figure}

Let $P$ be a polyhedron that is the vertical extrusion of a simple orthogonal polygon, such
as that illustrated in Figure~\ref{fig:partition}a.
The algorithm
begins by slicing $P$ into slabs, which in this special case are all
blocks (cuboids),
using $y$-perpendicular planes through each vertex.
The dual graph is a tree, $T_U$, having a node for each band and an edge
between each pair of adjacent bands.
In this special case, all $z$-beams are degenerate, i.e., of zero $z$-height.
The root is selected arbitrarily from among
all bands with a rim of
minimum $y$ coordinate. For example, the
polyhedron in Figure~\ref{fig:partition}a is sliced into nine
blocks, with $b_1$ as the root and
its unfolding tree as shown in Figure~\ref{fig:partition}b.
%

The rim of the root band with the
smaller $y$ coordinate is its \emph{front rim}, and the other rim
is its \emph{back rim}. For any other band, the rim adjacent to its parent in $T_U$ is
its front rim, and its other rim is its back rim.
Children attached along the front rim of their parent are \emph{front children}; children attached
along the back rim of their parent are \emph{back children}.
Note that ``front'' and ``back'' modifiers for rims and children derive from the structure of
$T_U$, and are not related to the ``forward'' and ``rearward'' $\pm y$
directions.
For example, $b_9$ is a front-child of $b_8$, although it is attached
to
the rearward face of $b_8$, and the front rim of $b_5$ lies
on the rearward face of $b_5$.

\subsection{Recursive Unfolding}

The key to the epsilon-unfolding method is the existence of a thin, non-crossing 
\emph{spiral} $\xi$ that cycles around each band
at least once, and unfolds to a staircase when flattened into the
plane.
A \emph{staircase} is an orthogonal path in the plane whose turns
alternate between $90^\circ$ left and  $90^\circ$ right, and so is a
monotone path.
The path that $\xi$ follows is determined recursively.
%
We review this \emph{spiral} 
$\xi$, starting with the base case.

\subsubsection{Single Band Base Case}
\label{sec:basecase}

\begin{figure}[htbp]
\centering
\includegraphics[width=\linewidth]{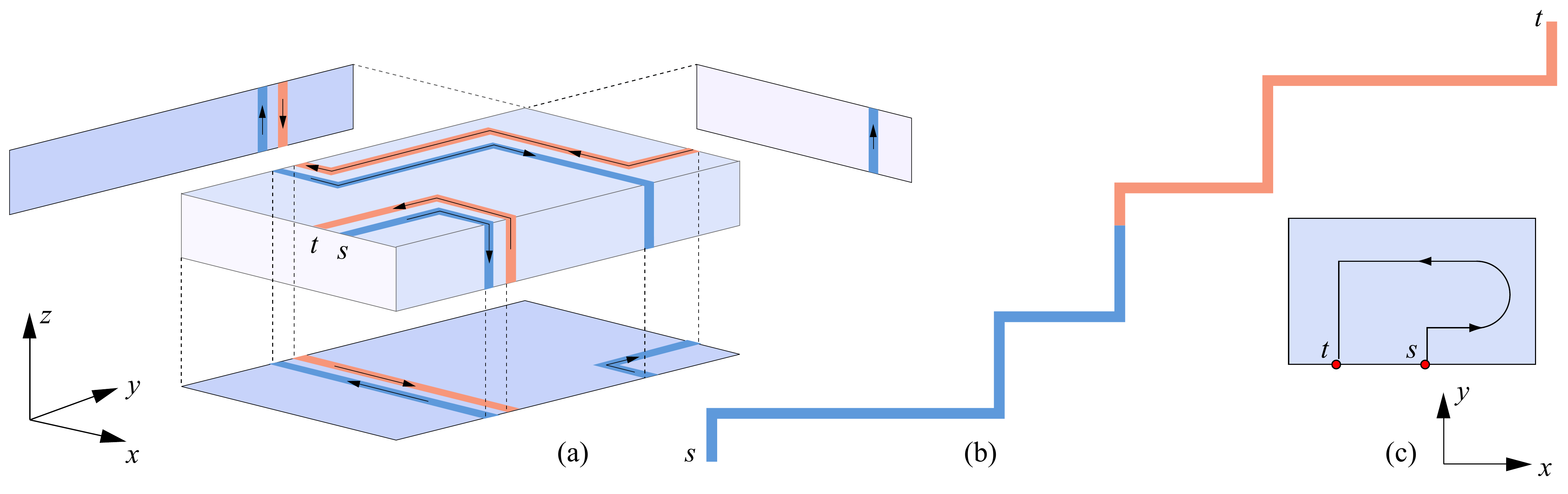}
\caption{(a) $R_{ts}$ block spiral, with
mirror views of faces that cannot be seen directly.
(b) Unfolded spiral.
(c) Abstract 2D representation.}
\label{fig:Rts}
\end{figure}

Figure~\ref{fig:Rts}a shows the path followed by $\xi$ 
for a single band corresponding to a leaf of $T_U$.
It starts at an \emph{entering point} $s$ on the top edge of
the front rim and spirals in a clockwise direction 
around the top, right, bottom, and left band faces 
toward the back rim. 
We call this spiral piece up to the point it reaches the back rim the
\emph{entering spiral}.
When it reaches the back rim, $\xi$ 
crosses the rearward face upward toward the top face.
From there, it retraces the entering spiral in the opposite (counterclockwise)
direction toward an \emph{exiting point} $t$  lying next to $s$ on
the front rim.
When $\xi$ is cut out, unfolded, and laid horizontally in the plane,
it forms a monotonic staircase strip, as shown in
Figure~\ref{fig:Rts}b, because the turns alternate between left and
right, $90^\circ$ each.
Observe that
the $x,z$-parallel segments of $\xi$, corresponding to the cycling
clockwise and counterclockwise around the band, form the stair ``treads'';
the $y$-parallel segments of $\xi$ and the $z$-parallel strip from the rearward face
form the stair ``risers.''

Three-dimensional illustrations of $\xi$ like that in Figure~\ref{fig:Rts}a
are impractical for
more complex orthogonal shapes. To easily illustrate more complex unfoldings, we
use the 2D representation
depicted in Figure~\ref{fig:Rts}c.
Note that the 2D representation captures the direction of the entering spiral and
the relative position of $s$ and $t$. The arc connecting the entrance to the exit
symbolizes the reversal of the unfolding direction
using a rearward face strip.

\begin{figure}[htbp]
\centering
\includegraphics[width=0.8\linewidth]{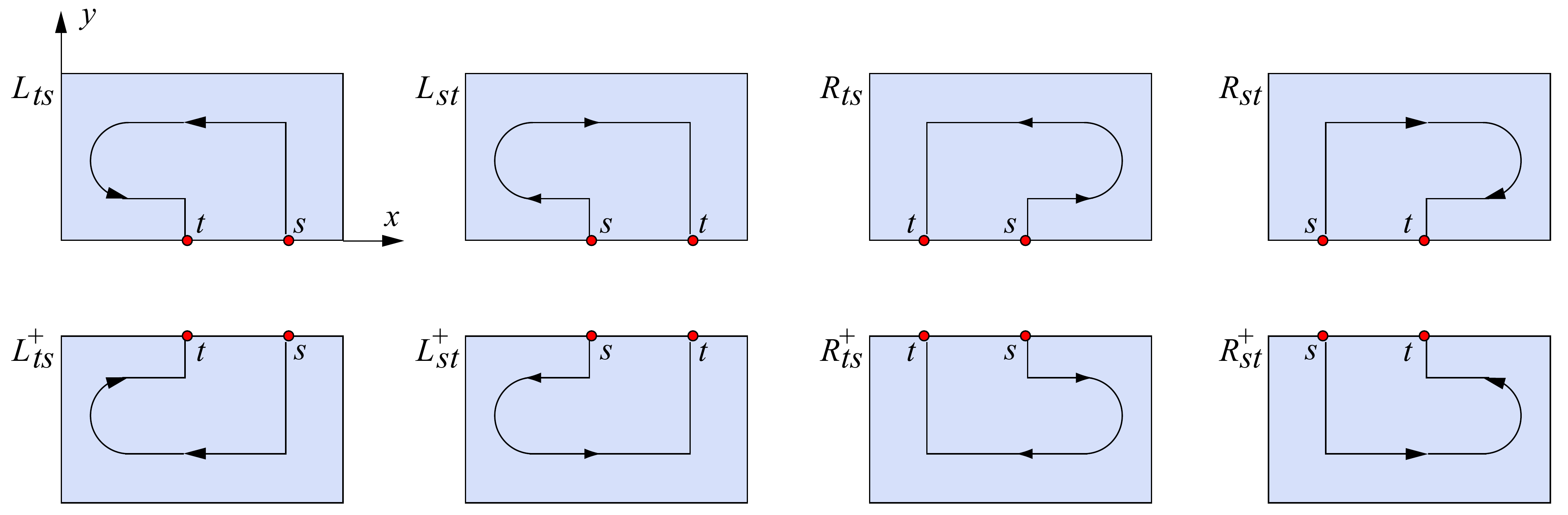}
\caption{Abstract 2D representations of the eight path types visiting one slab.}
\label{fig:R}
\end{figure}

Eight variations of the base case spiral are illustrated in Figure~\ref{fig:R}.
They differ in the manner in which $\xi$ enters and exits the band $b$ to be unfolded.
The four variations labeled $L_{ts}$, $L_{st}$, $R_{ts}$, $R_{st}$ in the top row are
used when the $y$-coordinate of $b$'s 
front rim is smaller than 
the $y$-coordinate of its back rim. $R_{st}$ is 
similar to $R_{ts}$, but 
with $s$ and $t$, and the clockwise/counterclockwise
cycling direction reversed; $L_{ts}$ and $L_{st}$ are (respectively) mirrors of
$R_{st}$ and $R_{ts}$ in an $x$-perpendicular plane. Note that the $R$ and $L$ labels indicate
the spiral's cycling direction when it enters the band: $R$ is clockwise,
$L$ is counterclockwise. The spiral exits the band cycling in the opposite direction.
The four variations in the bottom row
are labeled $L^{+}_{ts}, L^{+}_{st}, R^{+}_{ts}, R^{+}_{st}$, and they are
used when
the $y$-coordinate of $b$'s front rim is greater than the $y$-coordinate
of its back rim.
They are exact reflections of $L_{ts}$, $L_{st}$, $R_{ts}$, and $R_{st}$, respectively, in
a $y$-perpendicular plane.
The mirror symmetries imply that the 3D spiral corresponding to each 2D
abstract representation can be easily derived from $R_{ts}$ configuration, illustrated in Figure~\ref{fig:Rts}.

\subsubsection{Recursive Path}
\label{sec:eps-recursive}

For a node $b$ in $T_U$ that is not a leaf, we describe the path that the spiral $\xi$
recursively follows when visiting $b$.
We assume that $b$ has one of the eight configuration
labels shown in Figure~\ref{fig:R}.
As in the base cases, the label
identifies the relative order of points $s$ and $t$ on $b$'s front rim, the spiral's direction
when entering 
$b$, and 
$b$'s rim 
of lower $y$ coordinate.
Without loss of generality, we assume that
$b$'s label is $R_{ts}$; the other seven labels are equivalent by symmetry.
The inductive assumption is
that, for any subtree 
shorter than the subtree of $T_U$ rooted at $b$, and for
any configuration label assigned to the root band of the shorter subtree,
there is a (non-crossing) path $\xi$ consistent with that label
that cycles around each band in the smaller subtree at least once,
and unfolds in the plane as a staircase strip.

\begin{figure}[htbp]
\centering
\includegraphics[width=0.9\linewidth]{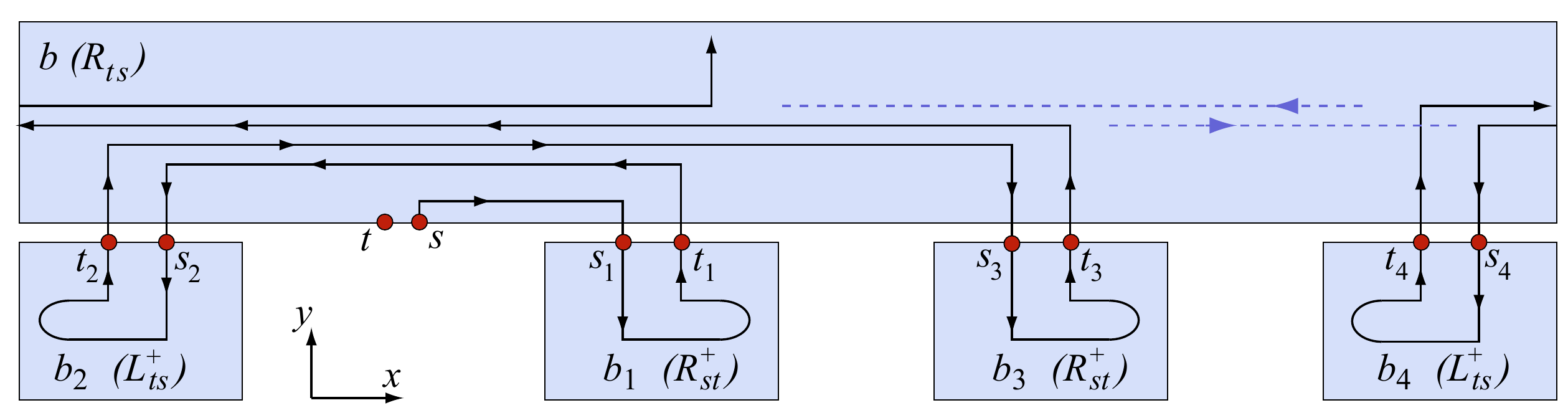}
\caption{Nested inside-out alternating path visits the front children. Dotted lines show $\xi$
where it cycles underneath on the bottom face.}
\label{fig:epsfrontchildren}
\end{figure}

After $\xi$ enters $b$ at point $s$,
it visits each of
$b$'s front children,
starting with the front child, call it $b_1$,
first encountered as it cycles clockwise along the front rim of $b$.
(See Figure~\ref{fig:epsfrontchildren}).
For reasons soon to be explained,
child $b_1$ is assigned the label $R^{+}_{st}$ with two points
$s_1$ and $t_1$ identified on the top edge of its front rim, with $t_1$
right of $s_1$.
The spiral $\xi$
enters $b_1$ at point $s_1$ and recursively visits it (and the subtree
it roots).
By the inductive hypothesis, $\xi$ exits $b_1$ at point $t_1$ cycling counterclockwise.
The label $R^{+}_{st}$ is assigned to $b_1$ because
$\xi$ is cycling in the direction $R$ (to the right, or clockwise)
on $b$ just before it enters $b_1$, and so it enters
$b_1$ with that same direction; the $+$ superscript is necessary
because the $y$-coordinate of the front rim of $b_1$ is 
higher than
that of its back rim; and the
$\nothing_{st}$ ordering is necessary to prevent $\xi$ from being
trapped beneath the portion of $\xi$ between $s$ and $s_1$ upon
returning to $b$, thus cutting itself off from reaching $b$'s other children
(because it cannot cross itself).

After recursively visiting $b_1$, $\xi$ 
cycles counterclockwise on $b$
to the first unvisited child it passes when on $b$'s top face. This child, call it $b_2$,
is assigned the label $L^{+}_{ts}$ with identified points $t_2$ and $s_2$ on its front
rim, consistent with its label,  and it is recursively visited.
In this nested manner, $\xi$ visits the
children clockwise and counterclockwise from $s$ from the inside out,
alternately assigning the labels $R^{+}_{st}$ and $L^{+}_{ts}$.
Figure~\ref{fig:epsfrontchildren} illustrates the path $\xi$ takes
when $b$ has four
front children. (To keep the example simple, only
one level of recursion is illustrated, with all children leaves of $T_U$.)

\begin{figure}[htbp]
\centering
\includegraphics[width=0.9\linewidth]{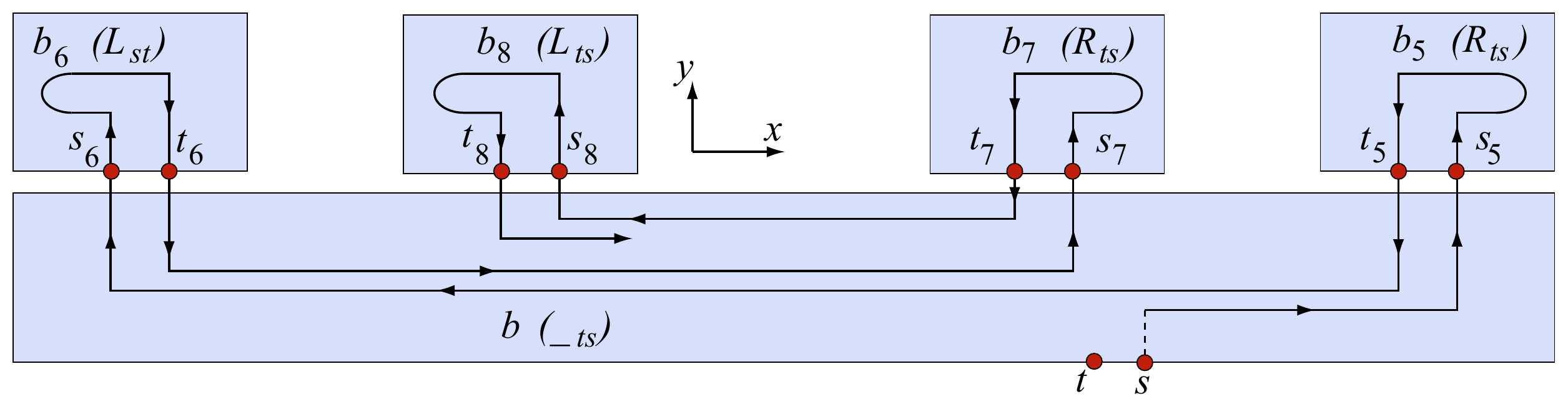}
\caption{Nested outside-in alternating path visits the back children.}
\label{fig:epsbackchildren}
\end{figure}

After visiting the front children, $\xi$ makes a complete cycle around $b$ and
then begins visiting the back children. Assume for concreteness that after visiting the
front children, $\xi$ is cycling 
clockwise on $b$, as shown in Figure~\ref{fig:epsbackchildren}. It then travels 
clockwise to the back child farthest to the right
along the top face. (See child $b_5$ in Figure~\ref{fig:epsbackchildren}).
When it returns from recursively visiting this back child, it will be cycling counterclockwise
(by the inductive hypothesis).
It is thus important that the child's exit point be to the left
of its entering point so that $\xi$ is not blocked from visiting other back children.
(See points $s_5$ and $t_5$ in Figure~\ref{fig:epsbackchildren}.)
Thus this first back child is assigned the configuration 
$R_{ts}$ and is recursively visited.  The spiral then moves to the unvisited child
farthest to the left
(see child $b_6$) and visits it in a similar way, assigning it the label $L_{st}$.
Thus the nesting of $\xi$'s alternating path is outside-in for back children,
with the labels $R_{ts}$ and $L_{st}$ being alternately assigned.

The last back child visited, $b_k$, however, is an exception when it comes to
its label assignment, for the following reason.
When the spiral exits $b_k$ (see $b_{8}$ in Figure~\ref{fig:epsbackchildren}),
it will retrace its path (in reverse direction) back
to the front rim of $b$ and then exit at point $t$.
For band $b$,  define its \emph{entering spiral}, $\xi_e(b)$, to be the portion of $\xi$ that begins
at $s$ and ends at the exiting point $t_k$ of $b_k$
($t_{8}$ in Figure~\ref{fig:epsbackchildren}).
Its \emph{exiting spiral}, $\xi_x(b)$, is the portion of $\xi$ that
begins at $t_k$
and ends at $t$ on the front rim of $b$.
The exiting spiral $\xi_x(b)$ simply parallels alongside the
entering spiral $\xi_e(b)$,
retracing the portion of $\xi_e(b)$ from $s$ to $s_k$ but in the opposite direction.
Since $b$ has a $\nothing_{ts}$ label, the exiting spiral must leave $b$ with the entering
spiral on its left, from the point of view of one walking on $b$
along the path taken by $\xi_x(b)$. Thus $b_k$
must also be assigned the label $\nothing_{ts}$ (consistent
with  $b$'s label), so that from the beginning of the retrace and throughout, $\xi_x(b)$ has
the entering spiral to its left. We call this a \emph{left retrace}; when $\xi_x(b)$ keeps
the entering spiral on its right during a retrace, we call it a \emph{right retrace}.
We note that if $b$ has no back children,
then the spiral reverses direction using a strip from $b$'s rearward face, as in
the base cases.

\subsection{Completing the Unfolding}
\label{sec:frontback}

We have focused on $\xi$'s recursive path because that is where the modifications
for delta-unfolding occur. But for completeness, we briefly summarize the
remainder of the epsilon-unfolding algorithm for extrusions, and
refer the reader to~\cite{Damian-Flatland-O'Rourke-2007-epsilon} for  additional details.
To complete the unfolding of $P$, $\xi$ is thicken in the $+y$ and $-y$ direction (as viewed
in the 3D coordinate system of Figure~\ref{fig:Rts}a) so that it completely covers
each band. This results in a thicker unfolded staircase strip. Then the
forward and rearward faces of $P$ are partitioned by imagining the band's top rim edges illuminating downward light rays in these
faces. The illuminated pieces are then ``hung" above and below the thickened staircase,
along the corresponding illuminating rim segments which lie along the horizontal edges of
the staircase.

\subsection{Level of Refinement}

\begin{figure}[htbp]
\centering
\includegraphics[width=\linewidth]{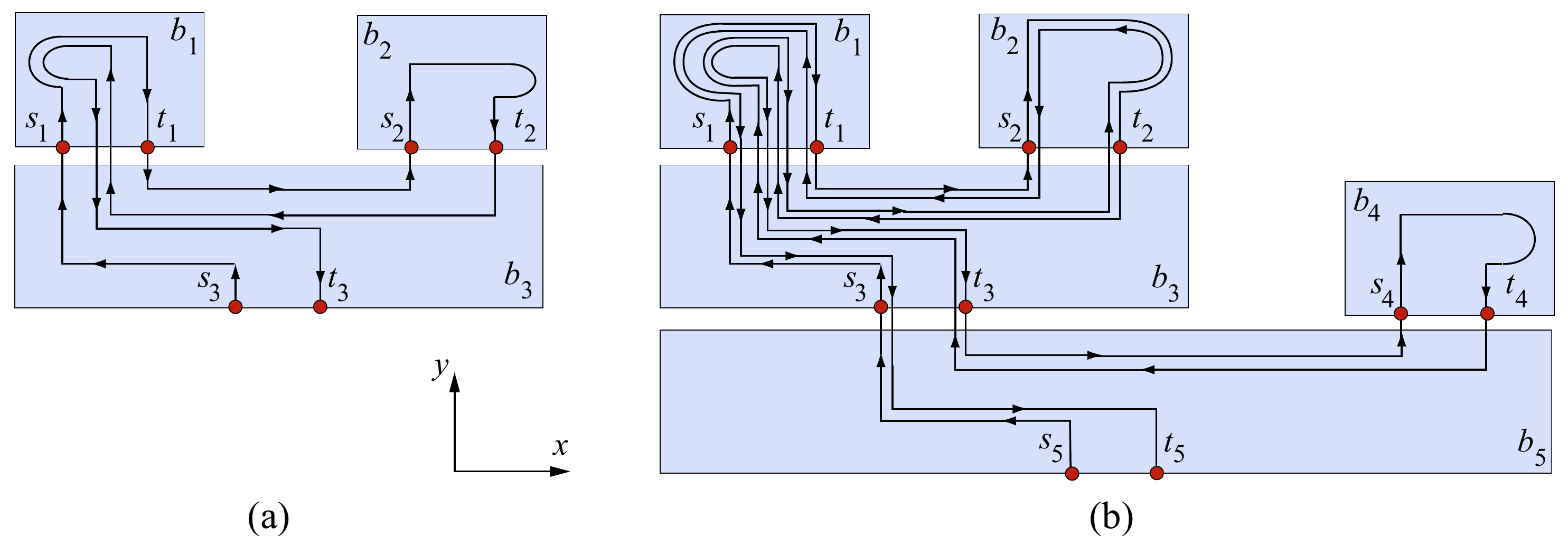}
\caption{Family of polyhedra requiring exponential refinement. Block $b_1$ is visited two times in (a), four times in (b), and in general $2^{\lfloor n/2 \rfloor}$ times for an $n$-block object.}
\label{fig:expcase}
\end{figure}

In~\cite{Damian-Flatland-O'Rourke-2007-epsilon} it was shown that the unfolding technique discussed so far 
can make an exponential
number of cuts on the family of polyhedra
depicted in Figure~\ref{fig:expcase}. Each polyhedron consists of $n =
2k+1$, $k \geq 1$, blocks
arranged as shown for $k = 1$ in Figure~\ref{fig:expcase}a, and for $k=2$ in Figure~\ref{fig:expcase}b. For analysis purposes, we formally define a \emph{visit} to a band to begin
when the spiral crosses its front rim to enter the band (either the first time, or in a retrace) and
end when it crosses the front rim to exit the band.
In Figure~\ref{fig:expcase}a, $b_3$'s visit begins when $\xi$ enters it at
point $s_3$ cycling counterclockwise. The spiral visits back child
$b_1$ and then $b_2$. The visit of $b_2$ triggers a retrace which involves a second visit of $b_1$, and then
back through $b_3$ to exiting point $t_3$, which completes $b_3$'s visit.
We can write this visit order using the string
$Q_3 = (s_3 ~(s_1~  t_1)~ (s_2 ~t_2)~ (s_1~ t_1)~ t_3)$, where an open parenthesis followed by a starting point
marks the start of a visit and  an exiting point followed by
a closing parenthesis marks the end.
The subscript on $Q$ is the number of blocks in the polyhedron. Observe that block $b_1$
is visited twice.
For the five block polyhedron in
Figure~\ref{fig:expcase}b, $\xi$ starts at point $s_5$ on $b_5$,
recursively visits block $b_3$ in the
manner just described, then visits $b_4$ which triggers a retrace through $b_3$.
After revisiting $b_3$, $\xi$ returns to $b_5$ and exits at point $t_5$. The corresponding visit string is
$Q_5 = (s_5~Q_3~ (s_4~t_4)~Q_3~ t_5) $.
The number of visits to $b_1$ doubles to $4$. In general,
an $n$ block polyhedron in this family gives rise to
$2^{\lfloor n/2 \rfloor}$ visits to $b_1$, resulting in an exponential
number of cuts on $b_1$.

\section{Delta-Unfolding Extrusions}
\label{sec:Delta}

To achieve quadratic refinement, we modify the
order in which children are visited
based on the heavy/light classification of nodes used in
\emph{heavy-path decomposition}~\cite{Sleator-Tarjan-1983}.
%
In heavy-path decomposition,
each tree node $v$ is
assigned a weight $n(v)$, which is the number of descendants in its
subtree, including itself.  An edge from parent $p$ to child $c$ is
\emph{heavy} if $n(c) > \frac{1}{2}n(p)$, and \emph{light} otherwise. We say a child
$c$ is \emph{heavy} (\emph{light}) if the edge between $c$ and its parent is heavy (light). Observe that
a node can have at most one heavy child.

If a node $b$ in $T_U$ has a heavy child, then we modify the
path of the entering spiral $\xi_e(b)$ so
that it visits the heavy child last, to prevent the need for revisiting the heavy child; we 
will show that this strategy quadratically bounds the
number of visits $\xi$ makes to each child.
For example, consider the polyhedron in Figure~\ref{fig:expcase}b, and observe that
$b_3$ is a heavy child. With epsilon-unfolding, $\xi_e(b)$ visited child $b_3$
before $b_4$. The visit to $b_4$ triggered a complete retrace of the subtree rooted
at $b_3$, thus leading to the visit string $Q_5 = (s_5~Q_3~ (s_4~t_4)~Q_3~ t_5)$,
and a total of four visits to $b_1$.
But if we reverse the visit order so that $\xi_e(b)$ visits $b_3$ after $b_4$,
then the visit
string becomes $Q'_5 = (s_5~(s_4~t_4)~Q_3~(s_4~t_4)~t_5)$, and no
block is visited more than twice.

Since any front or back child could be heavy, we focus first on the challenge of
finding a route for the entering spiral so that it visits any specified child
last.
If we can achieve this, then we can organize the visits to minimize
retracing.
We then
formally present the algorithm and
analyze the resulting level of refinement.

\subsection{Front Child Visited Last}
\label{sec:heavyfront}

\begin{figure}[htbp]
\centering
\includegraphics[width=0.8\linewidth]{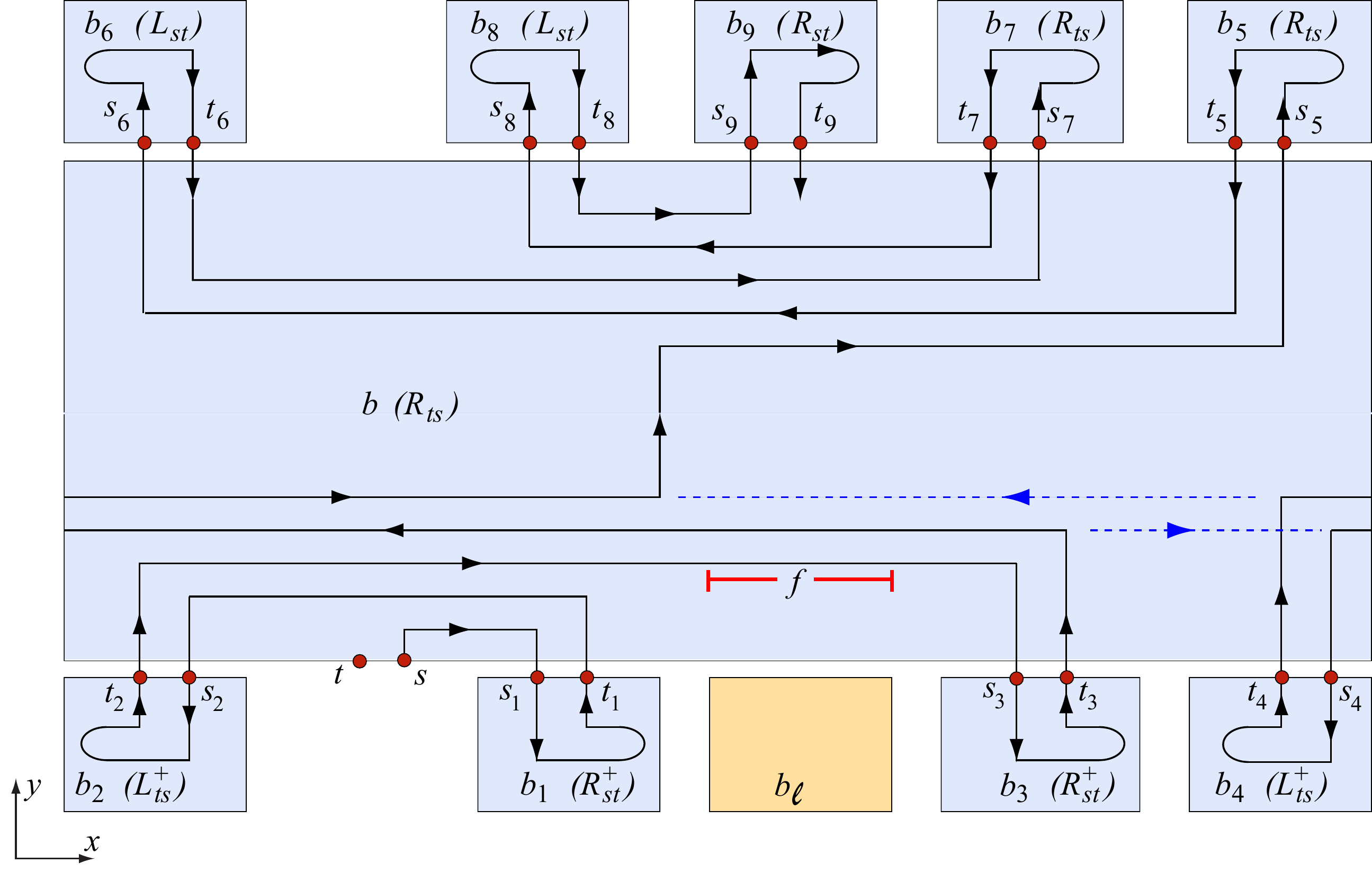}
\caption{Entering spiral  visits front and back children, with the exception of front child $b_\ell$,
which gets visited last (see Figure~\ref{fig:unfex2}).}
\label{fig:unfex}
\end{figure}

We start with the case when we desire to visit
a front child, call it $b_\ell$, last.
The idea is to visit all the front children excluding $b_\ell$,
and all the back children in exactly the 
manner described in Section~\ref{sec:eps-recursive}, as if $b_\ell$ were not present.
Figure~\ref{fig:unfex} shows the entering spiral $\xi_e(b)$ visiting all
but $b_\ell$.
(Note that the complete cycle that $\xi_e(b)$ makes between
 visiting the front and back children is not fully depicted in the
2D representation.)
After visiting the last back child ($b_9$ in the figure), $\xi$ retraces its
path in reverse.
It is during this retrace step that child $b_\ell$ is visited.
We explain the modifications necessary to accomplish this for a parent block
$b$ with a label of type $R_{\nothing}$ or $L_{\nothing}$; labels of type $R^+_{\nothing}$ and $L^+_{\nothing}$ are handled symmetrically.
%

\begin{figure}[htbp]
\centering
\includegraphics[width=0.8\linewidth]{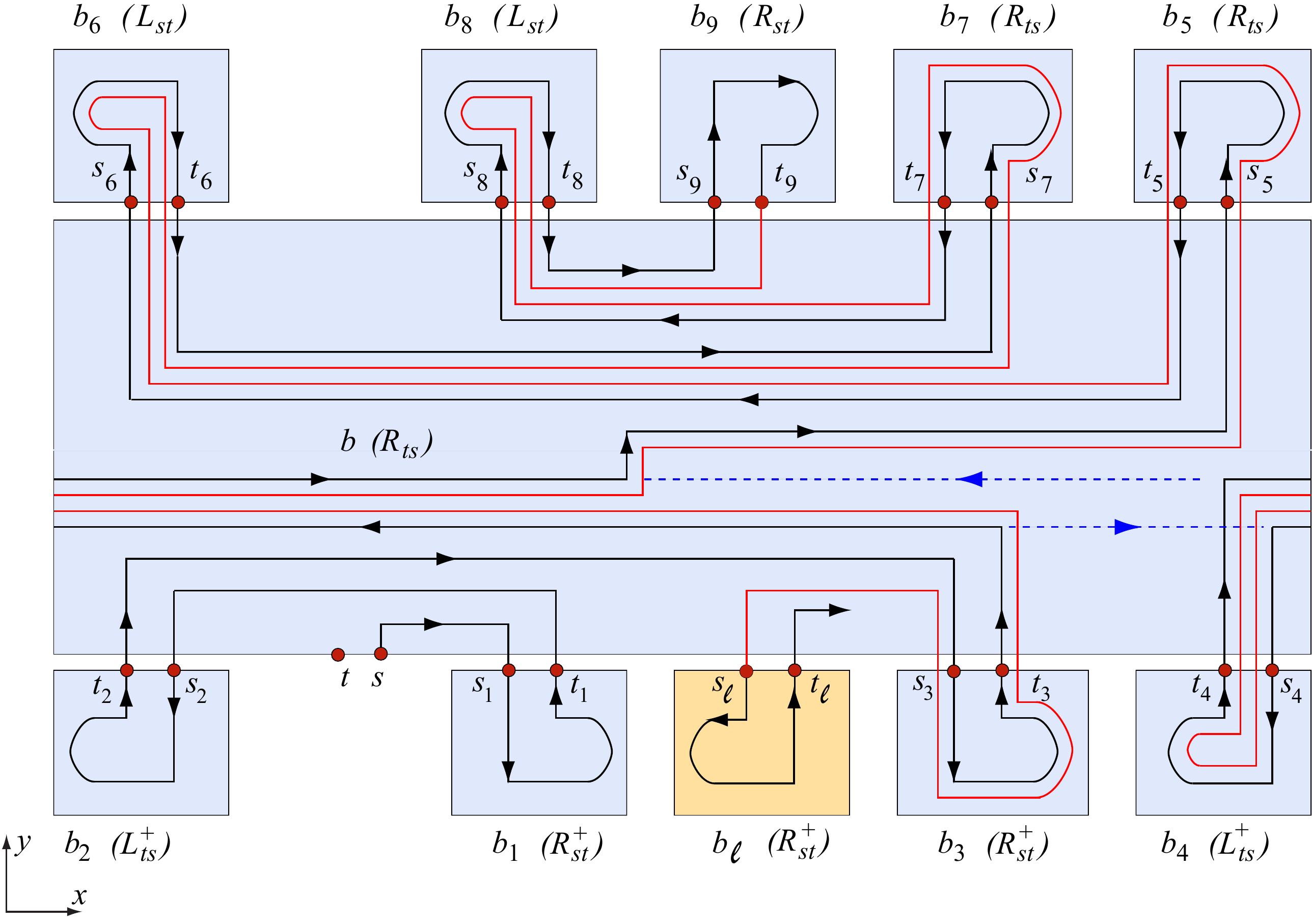}
\caption{Entering and return spirals. The return spiral passes by $b_\ell$ so that $b_\ell$
can be visited.}
\label{fig:unfex2}
\end{figure}

Observe first  that since $\xi_e(b)$ alternately visits all the front children
except for $b_\ell$ and then makes a complete cycle
around $b$,
some contiguous section of it, call it $f$, runs
alongside the top edge of $b_\ell$'s front rim.
Specifically, $f$ is the section of $\xi_e(b)$ hit by $y$-parallel rays shot from
$b_\ell$'s top front rim edge toward the back rim of $b$. See Figure~\ref{fig:unfex} where
$f$ is marked. Note that since
$\xi_e(b)$ is cycling toward the back rim of $b$, $f$ represents the first
time $\xi_e(b)$ 
passes by $b_\ell$'s top edge.
All subsequent passes are behind $f$.

During the retrace step,
 $\xi$ needs to run in front of $f$, so that it has unobstructed
access to $b_\ell$.
If the entering spiral
is cycling clockwise in section $f$, then the retracing spiral (which runs alongside
$f$ in the opposite direction) needs to
right-retrace, because that will keep the entering spiral to its right and
position it in front of $f$.
(Recall that clockwise is to the right and counterclockwise is
to the left.)
To trigger a right-retrace, we assign the last visited back child the label $\nothing_{st}$.
If the entering spiral
is cycling counterclockwise in section $f$, then the retracing spiral needs to
left-retrace, thus keeping the entering spiral on its left.
To trigger a left-retrace, we assign the last back child the label $\nothing_{ts}$.
So instead of matching the $\nothing_{st}$ or $\nothing_{ts}$ label of the last visited back
child to that of $b$ (as in epsilon-unfolding),
we instead assign it so that the retracing spiral passes alongside $b_\ell$.
When the retracing spiral reaches section $f$, it suspends
the retrace, enters $b_\ell$ at point $s_\ell$, and visits it. We call the portion of $\xi$
from the exiting point of the last visited back child to $s_\ell$ the
\emph{return spiral} and label it $\xi_r(b)$. See Figure~\ref{fig:unfex2}
which shows $\xi_r(b)$ in red extending from $t_9$ to $s_\ell$.

Upon exiting $b_\ell$ at point $t_\ell$, the spiral retraces its path
in the reverse direction, bringing it to the exit point $t$ on $b$.
Specifically, it follows the entire path from $s$ to $s_\ell$ in reverse.
This second retrace is $b$'s exiting spiral, $\xi_x(b)$.
(In Figure~\ref{fig:unfex2}, $\xi_x(b)$ is not illustrated, but
it begins at $t_\ell$ and follows the red and then the black path
to $t$, keeping them to its left.)
The label $\nothing_{st}$ or $\nothing_{ts}$ assigned to $b_\ell$ must be consistent
with 
$b$'s label in the following way. If $b$ has the label $\nothing_{ts}$, a left retrace
starting from $t_\ell$ is needed so that the spiral exits at $t$ on
the correct side of $s$ consistent with $b$'s $\nothing_{ts}$ label. Thus,
$b_\ell$ is assigned the label $\nothing_{st}$, the opposite of $b$'s label. If, however, $b$ has the label $\nothing_{st}$,
a right retrace is needed, and so $b_\ell$ is assigned the label $\nothing_{ts}$.

Because the label assigned to the last back child visited depends
on the direction of $\xi_e(b)$ in the $f$-section of the path, we show here that
determining that direction is straightforward.
We discuss the case in which $\xi_e(b)$ enters $b$
cycling clockwise; the case when it is cycling counterclockwise
is symmetric. We also assume that there are at least two front children
(not including $b_\ell$) and
they are labeled $b_1$, $b_2$, $b_3, \dots$,
in the order in which they are visited along the alternating path (as in Figure~\ref{fig:unfex}).
Observe that, if $b_\ell$ is located between $s$ and $b_1$ (as viewed from above),
then $\xi_e(b)$ first passes by $b_\ell$'s top edge cycling clockwise, and
the same is true if it is located between $b_i$ and $b_{i+2}$,
for $i$ odd ($i \in \{1, 3, 5, \dots\}$). Thus in these cases, $f$ is traversed clockwise.
If $b_\ell$ is located between $s$ and $b_2$ or
between $b_i$ and $b_{i+2}$, for $i \in \{2,4,6,\dots \}$, then
$\xi_e(b)$ first passes by the top edge of $b_\ell$ cycling counterclockwise. Thus
in these cases $f$ is traversed counterclockwise.
If the top edge of $b_\ell$ is to the right of the last  odd
numbered child or to the left of the last even numbered child, then
$\xi_e(b)$ first passes over $b_\ell$ during its complete cycle around $b$.
During this cycle, $\xi_e(b)$ is heading
clockwise  if the last visited child was even and counterclockwise
if the last visited child was odd. Cases when there are fewer than two
front children are easily handled: if $b_\ell$ is the only front child,
or if it is located between $s$ and $b_1$, then then $f$ is traversed clockwise;
otherwise, $f$ is traversed counterclockwise.

\subsection{Back Child Visited Last}
\label{sec:heavyback}

In this section we discuss the situation in which we desire
to visit a particular back child $b_\ell$ last.
In this case, $\xi_e(b)$ visits the front children as described in
Section~\ref{sec:eps-recursive}. It then visits the back children as described
in Section~\ref{sec:eps-recursive} but with an altered visiting order.
We consider the case when $b$ has a $L_{\nothing}$ or $R_{\nothing}$ type configuration
label and the
entering spiral $\xi_e(b)$ is cycling counterclockwise after visiting the front children;
the other cases are symmetric.

\begin{figure}[htbp]
\centering
\includegraphics[width=0.8\linewidth]{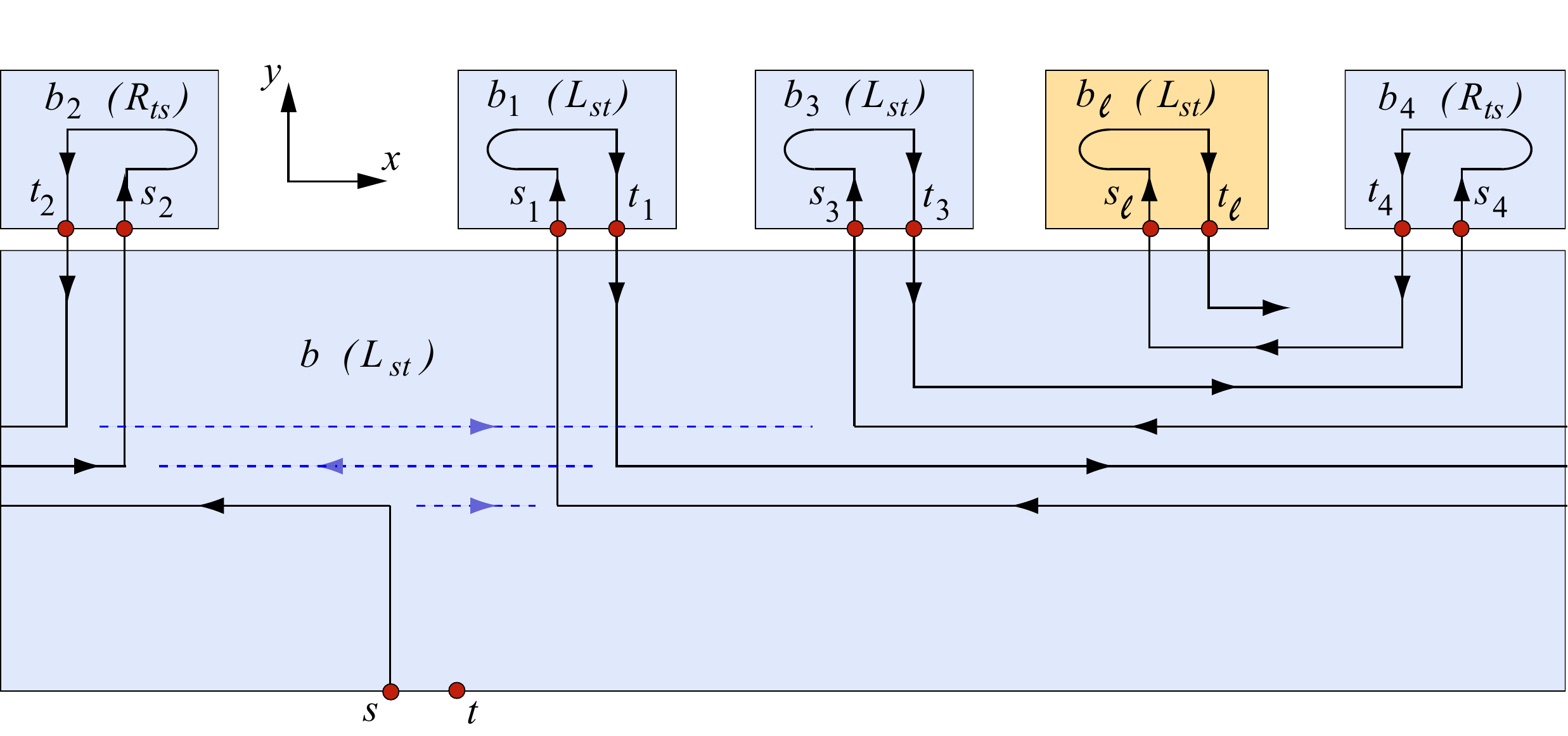}
\caption{Labels and path followed by spiral when visiting back children, when the last child
to be visited is back child $b_\ell$;
dashed
lines depict spiral pieces on the bottom of the parent block.}
\label{fig:unfback}
\end{figure}

Let $m \geq 0$ be the number of back children of $b$ not including $b_\ell$,
and let $b_1, b_2, \dots b_j$ be the front children, for $j \geq 0$.
Consider the back children of $b$ in the cyclic clockwise order in which their top edges
occur around $b$'s back rim.
When $m$ is odd, we label the $m$ back children
$(b_{j+1}, b_{j+3}, \dots, b_{m-2},  b_m, b_\ell, b_{m-1}, \dots,  b_{j+4}, b_{j+2})$,
according to their positions relative to $b_\ell$ in this cyclic ordering.
When $m$ is even, the labeling is 
$(b_{j+1}, b_{j+3}, \dots, b_{m-1}, b_\ell, b_{m}, \dots,  b_{j+4}, b_{j+2})$,
as depicted in Figure~\ref{fig:unfback}. The spiral
$\xi_e(b)$ visits the back children from the outside-in, following the visit order
$b_{j+1}, b_{j+2}, \dots, b_{m-1}, b_{m}, b_\ell$.
It is always possible to visit $b_{j+1}$ first,
with a full cycle of the spiral around $b$ (if necessary) to get the spiral
to the top edge of $b_{j+1}$. This is illustrated in Figure~\ref{fig:unfback}
for five back children (and no front children).

The assignment of $L_{\nothing}$ and $R_{\nothing}$ labels to the back children of $b$ is the same as described
in Section~\ref{sec:eps-recursive}.
Specifically, the labels for the children alternate between $L_{st}$
and $R_{ts}$ with respect to the visiting order.  The spiral $\xi_e(b)$ is cycling
counterclockwise (to the left) when it reaches $b_{j+1}$, which
matches $b_{j+1}$'s $L_{\nothing}$ label.
The recursive unfolding of $b_{j+1}$ reverses the direction of the spiral, so
that it enters
$b_{j+2}$
cycling clockwise (to the right), thus matching
$b_{j+2}$'s $R_{\nothing}$ label,
and similarly for the other back children.  The
alternating $\nothing_{st}$, $\nothing_{ts}$ labels of the children ensures an outside-in
nesting of $\xi$, which enables 
it to reach each back
child.  As in Section~\ref{sec:eps-recursive}, the one exception to the
alternating labels is the last visited child $b_\ell$, whose $\nothing_{st}$ or $\nothing_{ts}$ label
needs to match that of its parent $b$.
After visiting $b_\ell$, the exiting spiral follows the entering spiral in reverse to $t$
as in Section~\ref{sec:eps-recursive}, 
thus completing the visit of $b$.

\subsection{The Delta-Unfolding Algorithm for Extrusions}
\label{sec:analysis}

What we call the {\em delta-unfolding} algorithm is
a modified version of the epsilon-unfolding algorithm, 
which requires that at each node $b$ in $T_U$
with a heavy child, the spiral $\xi$ visits the heavy child last.
Specifically, if the heavy child is a front child, then $\xi$ follows the
path described in Section~\ref{sec:heavyfront}; if 
the heavy child is a back child, then $\xi$ follows the
path described in Section~\ref{sec:heavyback}.
If $b$ has no heavy child, then its children are visited in the
epsilon-unfolding order (Section~\ref{sec:eps-recursive}).
All remaining steps of the delta-unfolding algorithm for extrusions---the
thickening of $\xi$, the unfolding of $\xi$ as a staircase in the plane,
and the partitioning and hanging of the frontward and rearward faces from the
flattened staircase---are the same as for epsilon-unfolding.

\subsection{Refinement Analysis}

We now turn to analyzing the refinement for extrusions. The path taken by $\xi$
on a band is composed of a series of
axis-parallel segments.
We determine an asymptotic upper bound on the number
of such segments on any band face,
because this is an asymptotic upper bound on the total
number of cuts on a grid face in the unfolding. We compute this by bounding the number
of segments on any top face, as the number of segments on all four faces of a
band is asymptotically bounded by the number of segments on its top face.

Define the {\em first visit} of $\xi$ to a band $b$ to begin
when $\xi$ {\em first} enters $b$ at point $s$, includes the recursive visiting of
$b$'s children, and ends when it exits $b$ at point $t$.
Band
$b$ and the bands in its subtree may be revisited by $\xi$ many times during
subsequent retracings, but each of these retracings merely follows the path traced
during the first visit to $b$.
Let $R(n(b))$ be an asymptotic upper bound on the number of segments that
$\xi$'s first visit to $b$
induces on a top face of any band in the unfolding subtree rooted at $b$.
Then a bound on the number of
 segments on any top face in $b$'s subtree induced by $\xi$ (in its entirety) is $R(n(b))$
multiplied by the total number of times $\xi$ visits $b$.
We now establish three properties of $\xi$'s first visit to $b$:
\begin{enumerate}[(i)]
\item $\xi$ induces at most $O(n(b))$  segments on $b$'s top face;
\item the light children of $b$ are each visited at most four times; and
\item if $b$ has a heavy child, the heavy child is visited only once.
\end{enumerate}

For (i), the worst case occurs when
$b$ has $O(n(b))$ children and a heavy front child $b_\ell$.
In this case, the
alternating paths of $b$'s entering spiral $\xi_e(b)$
that have it visit each front child (excluding $b_\ell$) may induce
$O(n(b))$  segments on $b$'s top face, and similarly for the alternating paths to each
back child.
Then $b$'s return spiral $\xi_r(b)$ retraces these alternating paths up to the point
that it reaches $b_\ell$, which
at most doubles the number of  segments.
After visiting  $b_\ell$,
the exiting spiral $\xi_x(b)$ retraces the path $\xi_r(b)$ and then the path
$\xi_e(b)$ in reverse back to point $t$ on $b$, which again
at most doubles the number of  segments
on $b$. Thus the total number of  segments is $O(n(b))$.

For~(ii), the maximum visits to light children occur when
$b$ has a heavy front child. In this case,
$\xi_e(b)$ visits each light child once.
Then $\xi_r(b)$
visits each light child at most once on its way to the heavy front child. After visiting the heavy
front child, $\xi_x(b)$ retraces
$\xi_r(b)$
and then retraces
$\xi_e(b)$ to the entering point of $b$, thus visiting each light child
at most twice more. Therefore,
each light child is visited at most four times.

For~(iii), if $b$ has a heavy front child, then the path traversed by $\xi$
(detailed in Section~\ref{sec:heavyfront}) immediately establishes that the heavy front
child is visited only once. Similarly, if $b$ has a heavy back child, the path detailed
in Section~\ref{sec:heavyback} establishes that 
the heavy back child is visited exactly once.
%

Properties (i), (ii) and (iii) established above imply that $R(n(b))$ is determined
by the larger of three quantities:
\begin{enumerate}[(a)]
\item the number of  segments on $b$'s top face induced during $\xi$'s
first visit to $b$;
\item $4\max_{i=1\dots k} R(n(b_i))$, where $b_1, b_2,\dots b_k$ are $b$'s
light
children;
\item $R(n(b_\ell))$, where $b_\ell$ is $b$'s heavy child, if it has one.
\end{enumerate}
A multiplier of four is necessary in case~(b) because light children may be visited up to four times during $b$'s first visit; no multiplier is
necessary for the heavy child~(c) because it is visited only once. For the base case,
$R(1) = c$, for some constant $c > 1$, because the first visit of $\xi$ to
a leaf node band (as described in Section~\ref{sec:basecase}) induces a constant
number of  segments.
And in general,
\begin{eqnarray*}
 R(n(b)) & = & \max \left\{O(n(b)), ~4\max_{i=1\dots k} R(n(b_i)) , ~R(n(b_\ell))\right\} \\
            &  \leq & \max \left\{O(n(b), ~4\max_{i=1\dots k} R\left( {\textstyle \frac{1}{2}} n(b) \right), ~R(n(b)-1) \right\} \\
            &  = & \max \left\{O(n(b), ~4R\left( {\textstyle \frac{1}{2}}n(b) \right), ~R(n(b)-1) \right\}
\end{eqnarray*}
noting that the light children's subtrees contain at most $\frac{1}{2}n(b)$ nodes,
and the heavy child's subtree contains at most $n(b)-1$ nodes. It is
straightforward to verify by induction that $R(n(b)) = O(n(b)^2)$.
Applying this to the root $r$ of $T_U$ with $n = n(r)$ nodes
and noting that $\xi$ visits $r$ only once in the delta-unfolding
algorithm,
yields a maximum of $O(n^2)$  parallel segments on any top face.

This also bounds the number of cuts on any grid face in the unfolding.
Specifically, in the thickening step $\xi$ expands in the $+y$ and $-y$
direction so as to cover the entire band, but this does not asymptotically
increase its number of edges.
After the thickening, disjoint sections of
$\xi$ run along the entirety of both band rims. In the partitioning
step, the disjoint sections along the top rim edges
induce the division of the frontward and rearward faces into strips;
i.e., each disjoint section delimits the vertical strip beneath it. Because $O(n^2)$
bounds the number of disjoint sections along the top edge, it also bounds the number of strips
a frontward/rearward face is partitioned into.

\subsubsection{A worst case refinement example.}
A simple example establishes that the bound $O(n^2)$ is tight: a
polyhedron with $n = 2^{h+1}-1$ blocks,
whose unfolding tree $T_U$ is a perfect
binary tree of height $h$ (i.e., each internal node has two children, and all
leaves are at the same level).
There are no heavy nodes in $T_U$, and the number of cuts in a visit of
the root is given by the recurrence relation
$$
R(n)=4R((n-1)/2) = 4^{h}R(1) = (n+1)^2R(1)/4
\;,
$$
because
$$
4^h = 4^{\log_2(n+1) - 1} = 2^{\log_2(n+1)^2} /4 = (n+1)^2/4
\;.
$$
And since $R(1) = c$, for some constant $c$, it follows that $R(n) =
O(n^2)$,
establishing our claim.

\section{Delta-Unfolding of Genus-Zero Orthogonal Polyhedra}
\label{Delta-Unfolding of Genus-Zero Orthogonal Polyhedra}

The delta-unfolding algorithm and its refinement analysis
generalizes to all genus-zero orthogonal polyhedra in the same
way the epsilon-unfolding algorithm does, so we summarize the idea here and refer the
reader to~\cite{Damian-Flatland-O'Rourke-2007-epsilon} for details. Instead of partitioning $P$ into blocks, the general algorithm
partitions $P$ into slabs as defined in Section~\ref{sec:overview}.
It then creates an unfolding tree, $T_U$, where each node corresponds to a band surrounding a slab.
Each parent-child arc in $T_U$ corresponds to a $z$-beam, which is a vertical strip from a frontward or rearward face connecting the parent's rim to the child's rim.
For a parent band $b$, its front (back) children are those
whose $z$-beams connect to $b$'s front (back) rim.

\begin{figure}[htbp]
\centering
\includegraphics[width=0.8\linewidth]{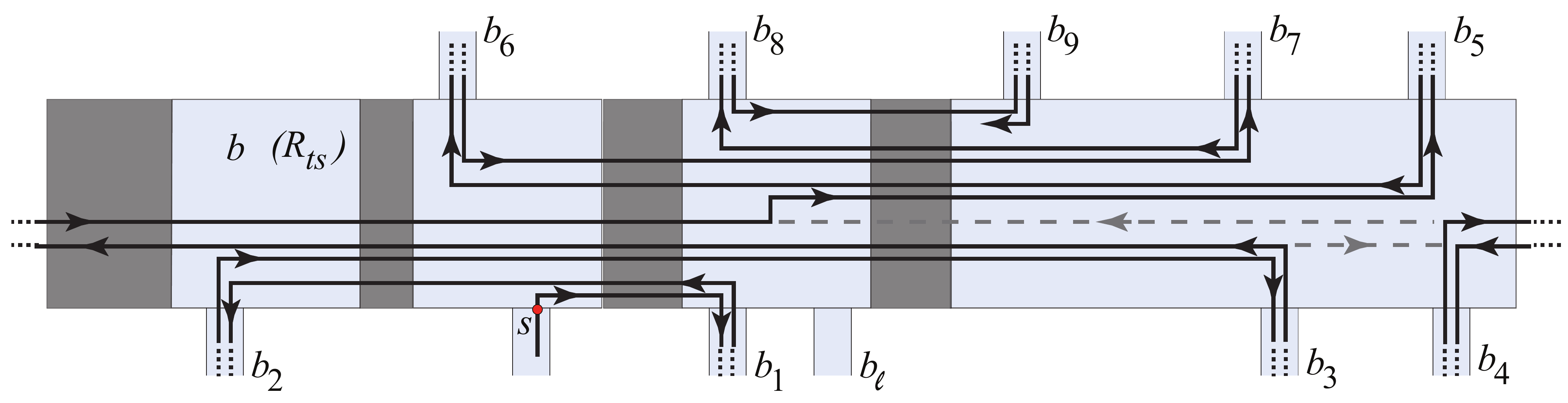}
\caption{Band $b$ of a slab, cut and laid flat with top/bottom faces light gray and right/left
faces dark gray. $z$-beam connections to $b$'s parent, children $b_1 \dots b_9$, and child $b_\ell$
are marked along the front and back rim. The path that $\xi_e(b)$ follows
when $b_\ell$ is heavy is depicted.}
\label{fig:genUnfEx}
\end{figure}

The spiral $\xi$ enters and exits $b$ at points $s$ and $t$
located at the intersection between $b$'s front rim and the $z$-beam
connecting $b$ to its parent.
Observe that there is a natural cyclic ordering of $b$'s front (back) children that is determined by their
$z$-beam connections around $b$'s front (back) rim.  Using this cyclic ordering, it
is straightforward to generalize the paths that $\xi$ follows to reach the
front and back children, described in Sections~\ref{sec:heavyfront} and~\ref{sec:heavyback}.
See for example Figure~\ref{fig:genUnfEx} that shows a band 
with its faces flattened in the plane (the lighter color marks top/bottom faces, and
the darker color marks right/left faces). Also depicted are
the $z$-beam connections (flattened into the plane) and the path $\xi_e(b)$  follows
to visit the children, assuming $b_\ell$ is a heavy child. Observe that the path is
the same as in Figure~\ref{fig:unfex}, except that it extends across multiple band faces.
When $\xi$ visits a child, it moves from $b$ to
the connecting $z$-beam and travels vertically (in 3D) along the $z$-beam to reach the child;
when it exits the child it travels along the $z$-beam back to $b$.
In the unfolded staircase, the portion
of $\xi$ on the $z$-beam corresponds to a vertical riser. Thickening $\xi$ is done
as in the case of 
extrusions. The partitioning of the
forwards and rearwards faces is also 
done as in the case of extrusions, but
in addition to shooting illuminating rays down from top rim edges, bottom rim edges that are not hit by these rays must
themselves shoot rays upward to illuminate portions of faces not illuminated by
the top edges. 
The face pieces resulting from this partitioning method
are hung from the staircase as described in Section~\ref{sec:frontback}. 

The $O(n^2)$ upper bound on the level of grid refinement for extrusions also applies to general orthogonal polyhedra by the following argument. In the case of extrusions, for any block $b$ with children, $\xi$ makes turns only
on $b$'s top face, because all access to the children is
from the top face; it makes no turns on the other three faces of $b$.
Therefore, we analyze the
number of segments (each corresponding to a turn) on the top face. For a band $b$
of an arbitrary orthogonal polyhedra, $\xi$ visits $b$'s children in the same manner as for
an extrusion, except that the turns made to access the children are made
on whatever top or bottom face has the connecting $z$-beam, as in Figure~\ref{fig:genUnfEx}. In particular, for a band $b$ with a given number
of front and back children, the same number of turns are made, whether
$b$ surrounds a block of an extrusion or a slab of a arbitrary orthogonal polyhedron.
In terms of maximum refinement, the worst case occurs when all the turns are
concentrated on a single face, which is exactly the situation handled by our upper bound
analysis in the case of extrusions.

\section{Conclusion}
We present modifications to the 
epsilon-unfolding algorithm from~\cite{Damian-Flatland-O'Rourke-2007-epsilon} that reduce the level of grid
refinement necessary to grid-unfold any
genus-zero orthogonal polyhedron from exponential to quadratic.
%
The next natural step is to seek a refined grid edge-unfolding of all genus-zero
orthogonal polyhedra that requires subquadratic refinement of the
grid faces, to date only achieved for highly restricted classes of
orthogonal polyhedra~\cite{Biedl-Demaine-Demaine-Lubiw-Overmars-O'Rourke-Robbins-Whitesides-1998,Damian-Flatland-O'Rourke-2008-manhattan,Damian-Meijer-2004-orthostacks}.
It is unlikely
that the technique used in this paper could be extended to produce such an unfolding, due
to the backtracking nature of our recursive unfolding algorithm. However, our preliminary
investigations embolden us
to conjecture that a constant refinement of the vertex grid
suffices to grid-unfold all orthogonal polyhedra.

\paragraph{Acknowledgement.} The authors would like to thank Joseph O'Rourke for his careful 
reading and helpful suggestions.

\bibliographystyle{alpha}
\bibliography{unfolding}

\end{document}